# Trailer Reimagined: An Innovative, Llm-DRiven, Expressive Automated Movie Summary framework (TRAILDREAMS)

**Roberto Balestri** [1*]
0009-0000-5008-2911

**Pasquale Cascarano** [1]
0000-0002-1475-2751

**Mirko Degli Esposti** [1]
0000-0003-0316-3449

**Guglielmo Pescatore** [1]
0000-0001-5206-6464

[1] Università di Bologna, Bologna, ITALY

[*] Corresponding author: roberto.balestri2@unibo.it

**ABSTRACT**

This paper introduces TRAILDREAMS, a framework that uses a large language model (LLM) to Accepted: 8 Mar 2025 automate the production of movie trailers. The purpose of LLM is to select key visual sequences and impactful dialogues, and to help TRAILDREAMS to generate audio elements such as music and voiceovers. The goal is to produce engaging and visually appealing trailers efficiently. In comparative evaluations, TRAILDREAMS surpasses current state-of-the-art trailer generation methods in viewer ratings. However, it still falls short when compared to real, human-crafted trailers. While TRAILDREAMS demonstrates significant promise and marks an advancement in automated creative processes, further improvements are necessary to bridge the quality gap with traditional trailers.

## INTRODUCTION

Movie trailers shape a film's market reception and box office success through brief, engaging narratives.

More than just promotional tools, trailers are crafted with a mix of artistic flair and strategic planning. This blend of creativity and strategy involves scene selection and narrative condensation, ensuring the trailer reflects the core themes of the film. Trailers' production typically represents around 5% of a movie's advertising budget, highlighting their significant role in film marketing (Wasko, 2003).

Since 2008, the evolution of trailer styles—from traditional voice-overs to a montage format with music— shows significant changes in production techniques. These shifts reflect a trend towards creating trailers that function as mini-narratives rather than straightforward advertisements, marking a new phase in film promotion (Richards, 2018).

Large language models (LLMs) like the generative pre-trained transformer (GPT) (Gallifant et al., 2024) series are advancing natural language generation by using extensive textual data to create coherent and contextually relevant texts. Historically, recurrent neural networks (RNNs) played a key role in processing sequential data, particularly in natural language tasks (Marhon et al., 2013; Piccolomini et al., 2019). However, RNNs faced challenges such as vanishing gradients and handling long-range dependencies (Tarwani & Edem, 2017). This led to a shift towards LLMs, which leverage transformer architectures and self-attention mechanisms, enabling them to comprehend and generate natural language with remarkable accuracy and fluency.

The integration of LLMs into various creative sectors is currently experimental but is already changing traditional approaches to content creation (Degli Esposti & Pescatore 2023). In music, literature, and visual arts, these models are being tested for their ability to generate creative content, assist in composition, and aid in editing and production. Artists and creators are exploring new forms of expression and innovation. This experimental use of LLMs is part of ongoing research to determine how these technologies can best support and enhance artistic creativity (Epstein et al., 2023).

Building upon a previous study conducted by the authors (Balestri et al., 2024a), this paper introduces TRAILDREAMS, a new framework that combines the capabilities of LLMs with other advanced AI technologies to automate the process of movie trailer production. This integration allows for the automated selection and assembly of key film scenes and dialogues, ensuring that trailers accurately reflect the film's storyline while engaging viewers.

TRAILDREAMS uses these technologies to thoroughly analyze a film's content, helping to identify essential narrative and genre-specific elements. This systematic approach ensures that trailers not only captivate potential viewers but also stay true to the film's narrative spirit.

To demonstrate the capabilities of this technology, we have developed hybrid trailers that merge traditional voice-over techniques with dynamic movie dialogues. These hybrids show how TRAILDREAMS can modernize trailer production by incorporating AI-driven analysis with conventional filmmaking techniques.

We first review the existing literature on automated trailer generation, categorizing it into three main methodologies. We also explore the application of LLM-driven approaches in video synthesis, highlighting their emerging significance. Then we provide a detailed explanation of TRAILDREAMS framework, outlining its components and how it operates. Next, we analyze the results generated by TRAILDREAMS, comparing them with established methods in the field, such as PPBVAM (Xu et al., 2015), MOVIE2TRAILER (Rehusevych & Firman, 2020), and traditional official trailers. We then present our critical analysis of the trailers generated, discussing potential improvements to various framework components that could enhance the quality of the outputs. Finally, we offer our concluding thoughts and reflections on the study.

## RELATED WORKS

While the area of video content summarization has been extensively investigated, the task of creating movie trailers automatically has received less attention. The prevailing body of work in video summarization predominantly aims to condense content, offering tools to manage the growing volume of video data rather than to craft engaging narratives as seen in movie trailers (Brachmann et al., 2009; Zhou et al., 2018).

The automatic generation of movie trailers is an interdisciplinary research area spanning natural language processing, computer vision, narrative extraction and multimedia content creation. Prior work has explored various techniques for trailer creation, though most have focused predominantly on visual and auditory aspects rather than holistic narrative approaches.

To group the studies by their approaches to automatic movie trailer generation, we can categorize them into three key methodologies:

**Visual and Auditory Feature Analysis**

Smeaton et al. (2006) developed a foundational method for creating movie trailers by selecting highlight clips based on visual activity, cinematography, and audio features, assembling them to form coherent trailers.

Zhou et al. (2010) advanced trailer generation with a focus on temporal visual features for scene categorization and classification, enhancing scene detection and genre classification but lacking in narrative integration.

Mahasseni et al. (2017) proposed a point process model with adversarial LSTM networks for trailer shot selection, concentrating on visual attractiveness and summarization, yet missing a multimodal approach that includes audio and textual analysis.

Rehusevych and Firman (2020) introduced "movie2trailer," employing anomaly detection to identify and compile unique, attention-grabbing frames into a trailer. While innovative in visual and audio feature extraction, their method does not fully address the integration of a comprehensive audio strategy, limiting the trailer's audio component to existing soundtrack segments without contextual alignment (Rehusevych & Firman, 2020).

**Emotion and Content Analysis**

Irie et al. (2010) introduced the VID2TRAILER system, utilizing affective content analysis to select emotionally impactful scenes and audio cues, considering factors like color, motion, and dialogue tone. The system also incorporates symbolic elements, enhancing thematic representation and brand recognition, to create trailers that effectively convey the film's mood and essence (Irie et al., 2010).

IBM research developed an AI-assisted trailer creation system that combines human and AI efforts. It leverages multimodal analysis to identify engaging clips, but the process requires human input for editing and narrative construction, highlighting a collaborative approach between AI capabilities and human creativity in trailer production (Smith et al., 2017).

**Narrative and Contextual Analysis**

Pavel et al. (2014) developed a video segmentation and textual summary generation method to improve content skimming and navigation. While this enhances viewer interaction with informational video content, it doesn't meet the specific needs of movie trailer creation, which requires a deeper integration of narrative, audio, and visual elements (Pavel et al., 2014).

Hesham et al.'s (2018) S-Trailer framework automates trailer generation using subtitle analysis to identify genre and key scenes, offering a text-centric approach to trailer production. However, this method might not fully capture the comprehensive visual and auditory aspects of traditional trailer creation (Hesham et al., 2018).

Papalampidi et al. (2021) introduced a graph-based movie representation for trailer generation, segmenting films based on screenplay-derived scene information. This innovative approach focuses on narrative progression and character dynamics but may be limited by the availability of screenplay data (Papalampidi et al., 2021).

Hu et al. (2022) utilized a graph convolutional network to segment films into meaningful parts based on narrative theory, Snyder (2005) aiming to highlight key narrative and emotional moments. While this method enhances scene representativeness, it incorporates a semi-automatic element due to manual segmentation (Hu et al., 2022).

Xie et al. (2023) presented a multimodal and aesthetic-guided approach for trailer creation, integrating various information types and aesthetic evaluation to ensure narrative and visual coherence. This method seeks to fully automate trailer production, enhancing both storytelling and aesthetic appeal (Xie et al., 2023).

**A Special Mention: LLM-Driven Video Generation**

In the realm of video generation with the use of LLMs, there are notable innovations that, while not directly aimed at automatic trailer generation, highlight the potential of LLMs in the broader context of video content creation.

For instance, Zhu et al.'s (2023) "MovieFactory" is a framework that transforms textual descriptions into complete movies. This system leverages ChatGPT to create detailed scripts, which are subsequently brought to life using advanced video and audio generation techniques (Zhu et al., 2023).

Similarly, Long et al.'s (2024) "VideoDrafter" framework utilizes LLMs to convert text prompts into detailed multi-scene scripts.

While these frameworks are not specifically designed for creating movie trailers, their methodologies highlight the versatility and power of LLMs in crafting engaging narrations.

In summary, existing literature has progressed from purely visual techniques to limited multimodal analysis for the movie trailer generation task. Our proposed TRAILDREAMS framework advances the field of trailer generation by integrating an LMM to drive a multimodal analysis, orchestrating audio, video, voiceovers, and dialogues from films. This software is distinct in its capability to manage and synthesize these diverse inputs. TRAILDREAMS ensures each element—from scene selection to audio creation—contributes to a trailer that offers an unprecedented level of narrative integration in automated trailer creation.

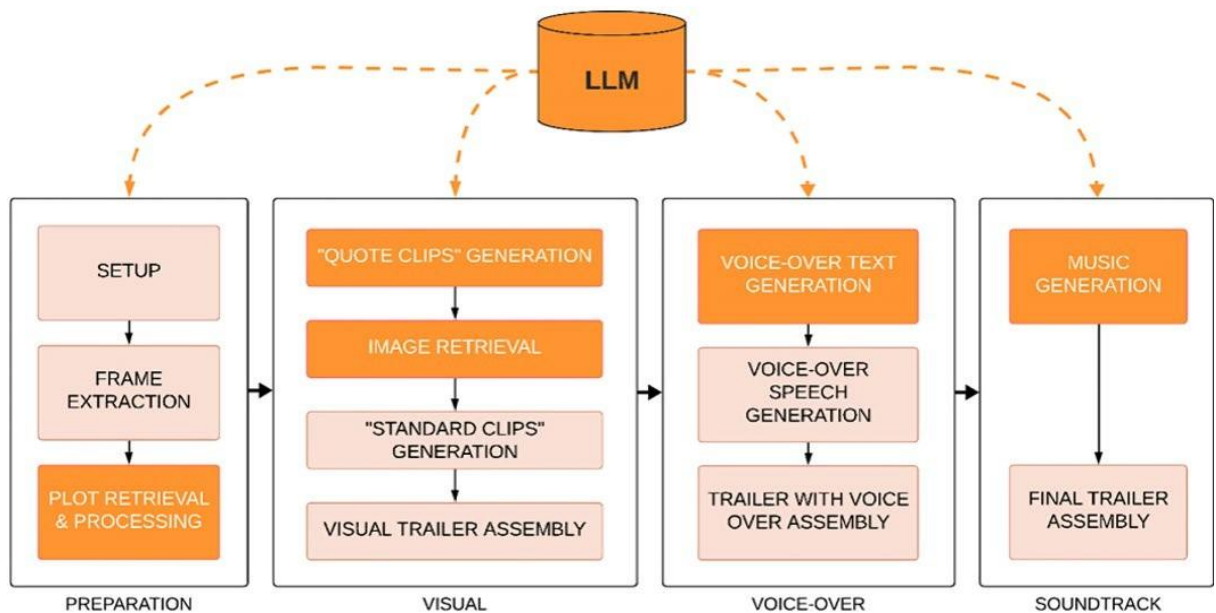

*Figure 1.* TRAILDREAMS framework process [Reprinted with permission from Balestri et al. (2024a). © IEEE]

# TRAILDREAMS FRAMEWORK

This section presents the proposed TRAILDREAMS, a framework for context-aware automated movie trailer generation.

**Figure 1** provides a conceptual map of TRAILDREAMS architecture, which is methodologically organized into four core stages:

1. **Preparation stage:** It involves initial setup, extracting frames from the movie, and dividing the movie synopsis into scenes using LLM.
2. **Visual stage:** It involves the creation of "Quote Clips (QC)," which are segments filled with impactful dialogues (selected by LLM) that capture the essence of the film's narrative, and "Standard Clips (SC)", which are the visual backbone of the story, guiding viewers through the unfolding plot without spoken words. These elements are then assembled into a coherent visual trailer
3. **Voice-over stage:** LLM generates a voice-over script that complements the visual content. This script is converted to audio, and the voice-over is synchronized with the visual trailer.
4. **Soundtrack stage:** A music generation model composes a unique soundtrack based that aligns with the film's themes thanks to LLM's musical direction. The final trailer is assembled by integrating this soundtrack with the voice-over and visual content, ensuring a balanced and impactful presentation.

LLM that "TRAILDREAMS" framework uses OpenAI's (2023a) GPT-4 interfaced via API calls. The framework's design is such that it could also be adapted to open-source models, suggesting that the approach is both innovative and accessible, with potential for broader application by researchers.

**Preparation Stage**

*Phase 0: Setup*

In the setup phase, the user can configure key settings to direct the movie trailer generation process. It includes specifying the unique identifier for the Internet movie database (IMDb) reference, the movie's file location and a project name for organizational purposes.

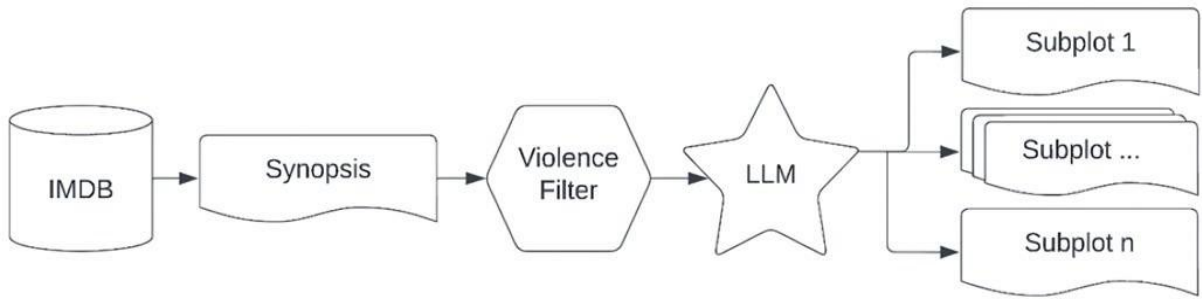

*Figure 2. The synopsis gets filtered and split into sub-plots (Source: Authors' own elaboration)*

IMDb code is needed as it enables the system to autonomously scrape and archive comprehensive movie information in preparation for subsequent processing steps. Utilizing the CINEMAGOER Python library, (Alberani, 2006) the system retrieves details such as the movie's synopsis, its most relevant quotes, release date, director(s), color technology, title, and genres. Additionally, a custom script specifically targets the movie's IMDb page to extract relevant quotes. This integration of IMDb information ensures that TRAILDREAMS is equipped with all the movie's data needed for the generation of the trailer.

The configuration also determines the desired quantity of SC and QC, although these targets are adaptable. The program has the flexibility to modify the final clip selection, possibly excluding some based on the overall trailer quality.

The user can set minimum and maximum lengths for the clips. Audio settings are adjusted to define appropriate volumes for voice-overs, SC audio, and background music, ensuring a balanced sound design.

*Phase 1: Frame extraction*

In the frame extraction phase, the process is designed to create a collection of frames from the film, which are needed for the subsequent trailer generation stages. The extraction is automatically performed by FFMPEG (Bellard, 2000) which is powerful and versatile open-source software designed to work with video and audio. Specifically, in phase 1, FFMPEG is utilized to achieve accurate frame captures at predetermined intervals.

This phase must reach an optimal balance in frame sampling. Sampling a few frames could limit the options available for matching various subplots of the movie, potentially omitting crucial visual moments. Conversely, excessive sampling may lead to redundancy and higher computational and storage costs.

To address this, the algorithm employs a strategic approach where the total count of frames to be extracted is determined by dividing the movie's duration by nine (so that one frame every nine seconds is extracted). This ratio is empirically derived, aimed at maximizing the representational diversity of the frames while ensuring a comprehensive coverage of the film's narrative arc.

The process commences by adjusting the movie's timeline to exclude the initial and final segments of the film, trying to avoid opening and ending credits and potential final spoilers. The algorithm then delineates timestamps for frame extraction, distributed uniformly across the adjusted timeline.

*Phase 2: Plot retrieval and processing*

In phase 2, LLM is employed to refine the movie's synopsis into a structured plot outline, serving as a foundation for the trailer's narrative framework. This critical process involves dissecting the core narrative elements to be accentuated in the trailer, ensuring that the essence of the movie is effectively communicated.

Initially, the synopsis undergoes a filtration process to eliminate any content that could trigger LLM's stringent content filters to prevent processing interruptions, such as violent, sexual or racist terms. For example, we filter out words like "killed", "cadaver", "corpse" substituting them with the word "REDACTED."

Subsequently, LLM segments the filtered synopsis into a series of "sub-plots." This step transcends basic summarization, as LLM creatively identifies and isolates key narrative elements that will enhance the trailer's narrative arc (**Figure 2**).

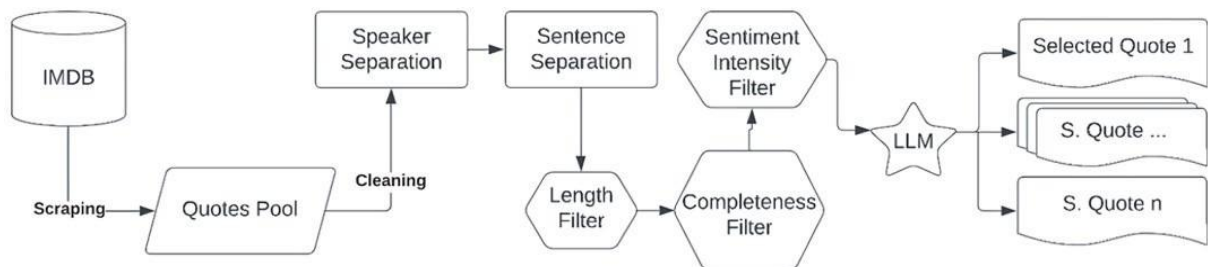

*Figure 3. Quote selection (Source: Authors' own elaboration)*

Each derived sub-plot is envisioned as a visual scene or sequence within the trailer, ensuring a compilation of significant, coherent, and engaging narrative fragments that, when combined, effectively convey the movie's storyline.

The resulting sub-plots are meticulously cataloged into individual folders, paving the way for a structured approach to the subsequent visual matching and trailer assembly phases.

For example, these are the first three subplots generated for Mission Impossible (De Palma, 1996) movie:

- *Agent Ethan Hunt in disguise preparing for a covert operation.*
- *Team of spies geared up in a dimly lit room, a sense of urgency in the air.*
- *Mysterious briefcase exchange gone awry amidst the Prague backdrop.*

**Visual Stage**

*Phase 3: Quote Clips generation*

In the "QC generation" phase, the system refines and selects key dialogues from the film to enhance the trailer's narrative depth. After gathering an initial collection of quotes, a thorough cleaning and filtering process is implemented to refine these quotes before they are analyzed by the language model.

The procedure starts with speaker separation, where the algorithm identifies and separates quotes from different speakers. Speaker separation specifically refers to organizing quotes pulled from IMDb, which initially appear grouped together like this into two isolated quotes attributable to Frodo and Gandalf individually, as follows:

> Frodo: "You're late."

> Gandalf: "A wizard is never late, Frodo Baggins. Nor is he early. He arrives precisely when he means to."

After cleaning the text by removing extraneous characters and spaces for uniformity, the system checks the length of each quote to ensure it adheres to predefined criteria (neither shorter than 12 characters nor exceeding 80 characters).

Employing the 'en_core_web_sm' SPACY (Explosion AI, 2016) NLP model, the system verifies the structural completeness of each sentence, confirming the presence of both a subject and a predicate. Sentences then undergo sentiment intensity analysis using the TextBlob Python library, ensuring each surpasses a minimal emotional weight threshold. These two steps are needed to filter out quotes that have a high probability of being irrelevant. Processed quotes are organized by length, prioritizing succinctness and emotional impact, with a cap to retain only the top 200 shortest quotes for the subsequent LLM evaluation.

The selected quotes are filtered from violent words and then passed to LLM, asking it to select those dialogues that better align with the film's core themes (**Figure 3**).

Subsequent steps involve audio extraction and transcription, utilizing StableWhisper (jianfch, 2023) a derivative of OpenAI's (2023b) Whisper speech recognition model, to transcribe audio. Given potential inaccuracies in speech recognition, a sequence matching algorithm inspired by Gestalt pattern matching (Ratcliff & Metzener, 1988) is employed to align selected quotes with corresponding audio segments accurately.

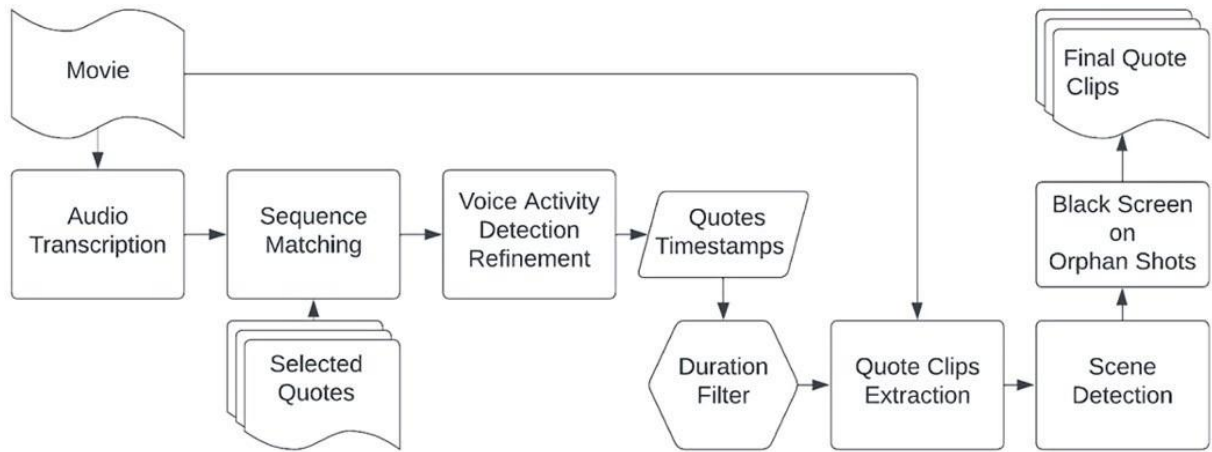

*Figure 4. Quote Clips generation (Source: Authors' own elaboration)*

To refine the timestamps of audio segments, enhancing their accuracy and reducing the likelihood of truncated phrases, a voice activity detection model, Pyannote (Bredin et al., 2020) is utilized. Intervals longer than 12 seconds are discarded to avoid very long QC (**Figure 4**).

The system then extracts video clips from the movie at the refined timestamps. A shot boundary detection algorithm (included in the PySceneDetect library) (Castellano, 2014) is applied to each clip to identify and rectify "orphan shots"—short, isolated shots that might appear erroneous in the final trailer. Detected orphan shots are replaced with black screens to maintain the trailer's visual continuity.

QC are unconnected to the subplots generated in phase 2; they are crucial because of their emotional impact rather than their narrative role.

For example, three quotes retrieved from the movie "Interstellar" (Nolan, 2014) are:

- *"Love is the one thing we're capable of perceiving that transcends time and space."*
- *"We've always defined ourselves by the ability to overcome the impossible."*
- *"Don't let me leave Murph."*

### Phase 4: Image retrieval

In this phase, inspired by previous work from Oliveira (2024) the system starts a process to align the film's narrative with corresponding visual representations. Keywords are extracted by LLM from the plot lines (the sub-plots we generated during phase 2). These keywords serve as anchors for the subsequent steps.

The core of this phase utilizes a multi-modal sentence transformer model, Clip-ViT-L-14 (Radford et al., 2021; Radford et al., 2022; Sentence Transformers, 2022) aimed to embed textual and visual content into a shared semantic space. This model, a derivative of sentence-BERT (Reimers & Gurevych, 2019) enables the system to understand and match the textual descriptors (keywords) with the visual content (movie frames), facilitating a semantic congruence between the two modalities.

Following the embedding process, the system shifts through the extensive collection of movie frames, now transformed into a structured array of embeddings. The objective is to identify those frames that exhibit a high degree of semantic similarity to the keywords. This similarity is quantified through cosine similarity measures, ensuring that the selected frames are semantically aligned with the narrative cues provided by the keywords. To ensure a representation of various parts of the movie, we check that the distance in seconds between diverse selected frames is at least 1.5% of the total duration of the movie (to avoid having, in the final trailer, shots from the same movie's scene).

To enhance the coherence of our trailers and minimize errors in the image retrieval process, we have adopted a selection method based on the temporal distribution of frames. While we recognize that a linear progression is not an inherent quality of human-made trailers, empirical evidence has shown that aligning the first 40% of the generated subplots with the initial 40% of the movie's frames, and the remaining scenes with the subsequent 60%, improves the quality of the generated trailers.

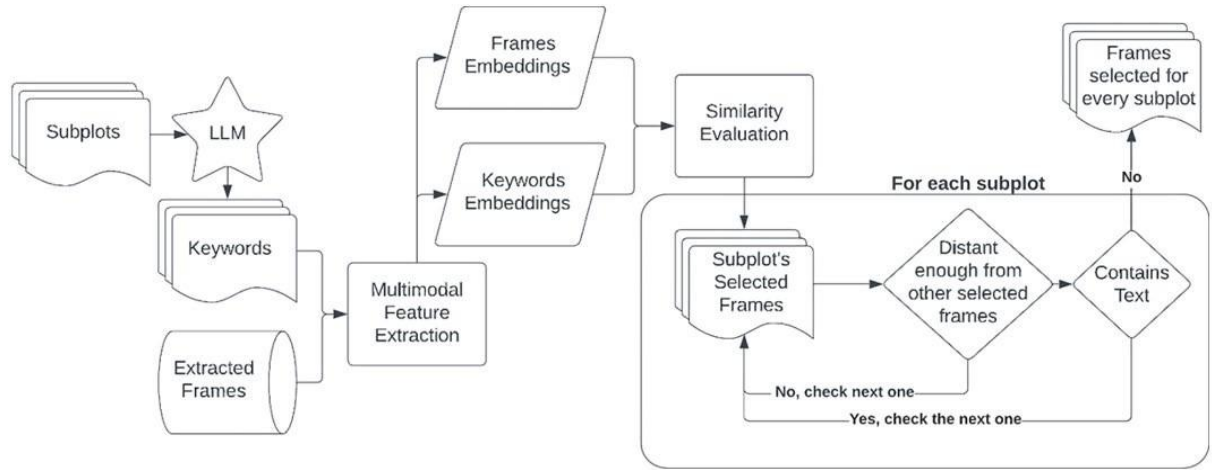

*Figure 5. Image retrieval process (Source: Authors' own elaboration)*

We also ensure, using the Python Library EASYOCR (JaidedAI, 2020) based on the CRAFT (Baek et al., 2019) algorithm and a CRNN (Shi et al., 2017) recognition model, that the selected frame does not contain superimposed text. Should text be detected, the system opts for the next most similar frame.

The frames selected for each scene are now devoid of extraneous text and semantically aligned with the movie's narrative (**Figure 5**).

*Phase 5: Standard Clips generation*

In this phase the system creates video segments starting from the selected frames of the previous phase.

Around each selected frame, the system establishes a "buffered zone", a temporal window that extends before and after the frame. This buffer ensures that the generated clip captures the full context and essence surrounding the key moment.

Within this buffered zone, the system employs shot boundary detection algorithms to identify a precise start point for the clip. This step ensures that each clip is a coherent and complete narrative unit, starting and ending at natural breaks in the video content, trying to keep the clips with a duration between the minimum and maximum length specified during the setup phase.

To further refine the clips, the system evaluates the consistency of the shots within each clip with a more aggressive shot boundary detection. It identifies and rectifies any "orphan shot" enhancing the clip's narrative flow and visual continuity. In this case the system does not overlay the orphan scenes with black screens like we did with QC but shorten or extend the clip to be visually appealing trying to maintain the minimum and maximum length specified before.

*Phase 6: Visual trailer assembly*

In this phase, the algorithm combines standard and QC to forge a coherent narrative flow. Initially, the system sorts standard scene clips aligning them in a sequence that reflects their original narrative order. This organization helps prevent inconsistencies that might arise from the image retrieval phase.

In the proposed method, QC are initially processed using a hybrid transformer model for audio source separation, as described by Rouard et al. (2023). This model isolates the vocal elements from the background, maintaining only the necessary vocal parts. Following this, the integration of these QC into the sequence of standard scene clips is managed through a systematic interval strategy. This strategy involves dividing the total number of scene clips by the sum of the number of QC plus one, thereby determining optimal insertion points for the QC within the sequence (see **Figure 6**). For example, if a trailer has 7 clips in total, of which 3 are QC and 4 are SC, we will have a sequence like this: SC, QC, SC, QC, SC, QC, and SC.

To ensure seamless auditory transitions, each QC is introduced with swift fade-in and fade-out effects, enhancing the smoothness of the auditory changes and minimizing the perceptual jarring between different audio clips.

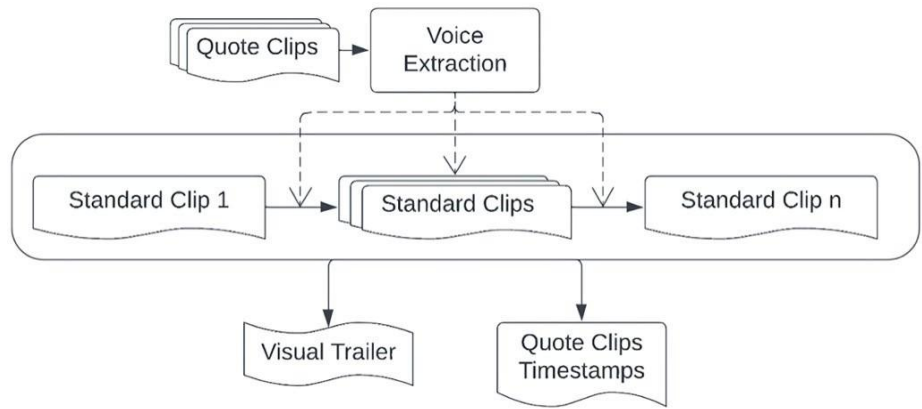

*Figure 6.* Visual trailer assembling (Source: Authors' own elaboration)

Finally, the system concatenates all clips into one unified video file and generates a timestamp log. This log records start and end times of each QC in generated trailer, providing information for later audio editing. **Voice-Over Stage**

*Phase 7: Voice-over text generation*

In this phase, LLM utilizes the movie's plot summary, directorial credits, and release date to generate a series of evocative and intriguing phrases suitable for the trailer's voice-over. Care is taken to ensure these phrases are engaging and lyrical while avoiding spoilers, thereby maintaining the film's suspense and allure. The quantity of phrases produced is tailored to match the length of the trailer constructed in the previous steps, ensuring a harmonious integration of narrative and visual elements. For example, the four phrases generated for the movie "The Hobbit – An Unexpected Journey" (Jackson, 2012) are:

- *"Every legend has a beginning, every adventure has an unexpected hero."*
- *"On the edge of the greatest quest, in the smallest of hands lies the fate of Middle Earth."*
- *"Whispered in shadows, an evil stirs waiting for the gold-touched fire."*
- *"Stand with Peter Jackson as destiny awakens in December."*

*Phase 8: Voice-over speech generation*

In this phase, the algorithm leverages the Coqui xtts-v2 text-to-speech (TTS) open-source model (Coqui AI, 2021) to convert scripted text into audio, utilizing an approach that aligns voice selection with the movie's genre(s). From a predefined set of five voice types sourced from the LibriTTS dataset (Zen et al., 2019) the algorithm selects a voice that aligns with the movie's genre.

For films that encompass multiple genres (it happens very often that a movie has various genres attributed to IMDb), the algorithm employs a genre-to-voice mapping strategy where each genre is associated with a specific voice type. When a movie is tagged with several genres, the algorithm counts the occurrences of each voice type corresponding to these genres. The voice type that is most frequently associated with the movie's genres is selected. Then the TTS model generates audio files for each voice-over script segment. From now on, we'll call these files "Voice Clips."

*Phase 9: Trailer with voice-over assembly*

In this phase, the algorithm integrates Voice Clips into the trailer, focusing on aligning them with the visual content and adjusting the volume of QC to maintain their balanced volumes. The algorithm initially retrieves the timestamps (that we saved before) that indicate the placement of QC in the trailer. Using these timestamps, the algorithm positions Voice Clips on the trailer's timeline, ensuring there is no overlap between them.

A key step involves calculating the average volume of the Voice Clips. This value is used to adjust the volume of QC, ensuring they integrate well into the trailer's audio. After adjusting the volumes, the algorithm combines the Voice Clips with the trailer's current audio track and reattaches the audio to the video.

**Soundtrack Stage**

*Phase 10: Music generation*

The algorithm executes a detailed process to synthesize background music for the trailer, ensuring the audio resonates with the story's mood and theme. Acting like an innovative composer, LLM crafts a detailed music description derived from the movie's plot. This description encapsulates the essence of the narrative, setting the stage for the music creation process. For example, the musical description generated by LLM for Mission Impossible (De Palma, 1996) is:

- *Instruments: Symphony Orchestra with a prominent use of brass and percussion, electronic elements for texture.*

- *Key: Varies, utilizing modal interchange.*
- *Tempo: Range from Andante to Allegro for different sections.*
- *Dynamics: Wide dynamic range, from pianissimo in covert scenes to fortissimo during action sequences.*
- *Texture: Complex, with motifs weaving between sections.*
- *Rhythmic Elements: Syncopation and mixed meter for creating urgency and tension.*
- *Harmonic Elements: Dissonant chords and chromatic movement for mystery and suspense.*
- *Melodic Elements: Leitmotifs for characters, using minor scales and octave leaps for intensity.*

The MusicGen (Copet et al., 2024) text-to-audio model is then prompted to generate music that corresponds with the text produced by LLM.

### Phase 11: Final trailer assembly

The system integrates the created music with the trailer's visuals and existing audio, employing audio ducking techniques to maintain audio balance. In fact, the algorithm lower the music volume when there are QC or Voice Clips, preserving narrative clarity. Post-ducking, the system combines the adjusted audio with the video. This step includes applying fade-in and fade-out effects to both audio and video, ensuring a seamless viewing experience.

### System specifications and efficiency

In the development and testing of TRAILDREAMS framework, our setup included an HP Omen 16 laptop, equipped with an Intel i7 13700HX processor, NVIDIA RTX 4080 graphics card, and 16GB RAM, running on the Windows 11 operating system. We utilized Python 3.10 for our programming needs. Notably, despite the advanced capabilities of our framework, TRAILDREAMS does not necessitate such high-end platforms for efficient functioning. This efficiency is exemplified in our case study with "The Wolverine," a movie with a runtime of 2 hours and 18 minutes. Utilizing our framework on the specified setup, the process to generate a trailer for this film was completed in 24 minutes. This underscores the framework's accessibility and its ability to deliver good performance even on mediocre systems.

## EVALUATION AND RESULTS

In this section, we present a comparative analysis, focusing specifically on TRAILDREAMS framework and leading competitors in automatic trailer generation. Our evaluation is based on the trailers produced by these methodologies, which are available for direct comparison. While we would have preferred to include trailers generated by more recent papers for a broader evaluation, we were unable to access them, limiting our comparison to the available data. The competitors initially included:

- VID2TRAILER (Irie et al., 2010),
- MUVEE – commercial software for video summarization,
- PPBVAM (point process-based visual attractiveness model) (Xu et al., 2015), • MOVIE2TRAILER (Rehusevych & Firman, 2020), and
- RT – The original official real trailers.

However, based on the results reported in Rehusevych and Firman (2020), which demonstrated that both VID2TRAILER and MUVEE trailed significantly behind in performance, we have excluded these two from our current analysis, limiting our competitors to PPBVAM, MOVIE2TRAILER, and the official real trailer.

To guarantee an unbiased assessment, we implemented a stringent methodology. We recruited 16 volunteers with diverse movie preferences (bachelor's and master's students in media studies) to review the trailers. To ensure impartiality, none of the participants had seen the trailers before, and they were not informed about the order in which the methods were applied. For consistency, all trailers were standardized to a resolution of 480x360. In contrast to other methods such as PPBVAM and MOVIE2TRAILER, which utilize the original movie soundtracks, our TRAILDREAMS framework stands out by incorporating automatically generated voiceovers and soundtracks. This approach underscores our dedication to delivering a fullyautomatic and immersive audio-visual experience. Additionally, while it is possible to run TRAILDREAMS multiple times to select the best output, the trailers we presented to the volunteers were generated from a single execution of our software. This approach might differ from that of our competitors, who may have chosen the best human-evaluated output from their framework. Similar to Irie et al. (2010), Xu et al. (2015), and Rehusevych and Firman (2020), volunteers assessed each trailer on appropriateness, attractiveness, and interest to determine the likelihood of watching the original movie after viewing the trailer:

- appropriateness: "How similar this trailer looks to an actual trailer?"

- attractiveness: "How attractive is this trailer?"
- interest: "How likely you are going to watch the original movie after watching this trailer?"

They provided ratings on a Likert (1932) scale from 1 (lowest) to 7 (highest). The analysis focused on trailers for "The Wolverine" (2013), "The Hobbit: The Desolation of Smaug" (2013), and "300: Rise of an Empire" (2014). These specific movies were chosen due to the available trailers from the other studies and constraints related to reproducing certain components of competitor algorithms.

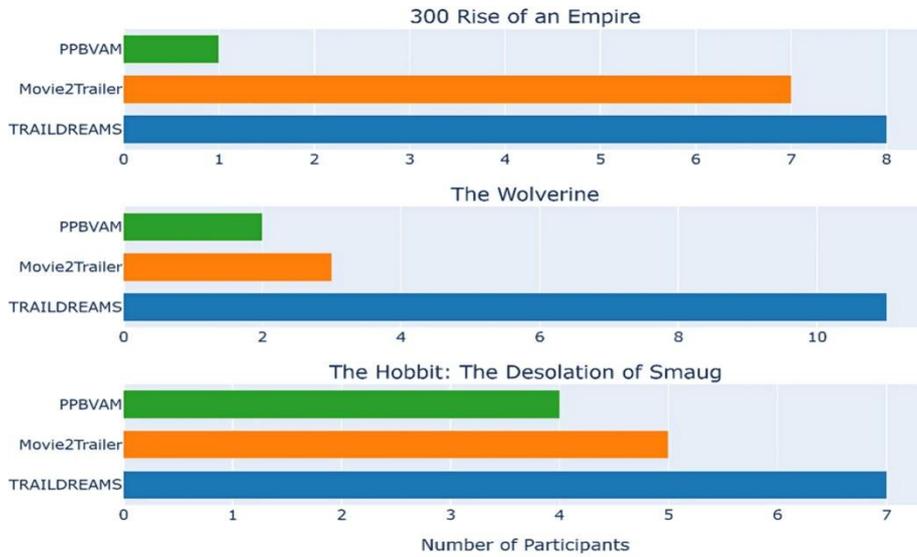

*Figure 7. Best trailer for movies for participants [Reprinted with permission from Balestri et al. (2024a). © IEEE]*

We evaluated the effectiveness of different trailer generation methods using a total score metric, which is the sum of the ratings across three metrics. For each participant, the method with the highest total score was deemed the most effective. **Figure 7** displays a bar plot showing the number of participants who rated each method as the best for the movies considered, offering a direct comparison of how well each method meets viewer criteria across various cinematic contexts. TRAILDREAMS consistently outperformed its competitors, indicating its superior ability to convince viewers. Notably, TRAILDREAMS' trailer for "The Wolverine" received high ratings, maybe due to the good quality of soundtrack and voiceovers generated for this movie compared to the ones generated for the other two movies.

Furthermore, we calculated the average and median scores for each method, presented in **Table 1** and **Table 2**. TRAILDREAMS demonstrated superior performance over PPBVAM in all categories and outperformed MOVIE2TRAILER in appropriateness and interest. The slight variation in attractiveness scores may be attributed mainly to the AI-generated soundtracks used by TRAILDREAMS that lack the quality of the original human-composed soundtracks used by competitors. Nevertheless, TRAILDREAMS' performance is commendable, particularly in maintaining competitiveness in Attractiveness along with strong results in Appropriateness. The latter underscores TRAILDREAMS' ability to produce trailers that meet traditional expectations for how movie trailers should look and feel.

*Table 1. Average scores by method across three movies for each quality metric.*

| Mean score | TRAILDREAMS | Movie2Trailer | PPBVAM |
|---|---|---|---|
| Appropriateness | **3.56** | 3.15 | 2.81 |
| Attractiveness | 2.96 | **3.08** | 2.88 |
| Interest | **2.90** | 2.85 | 2.62 |

*Table 2. Median scores by method across three movies for each quality metric.*

| Median score | TRAILDREAMS | Movie2Trailer | PPBVAM |
|---|---|---|---|
| Appropriateness | **4.0** | 3.0 | 2.0 |
| Attractiveness | 3.0 | **3.0** | **3.0** |
| Interest | 3.0 | 2.0 | 2.5 |

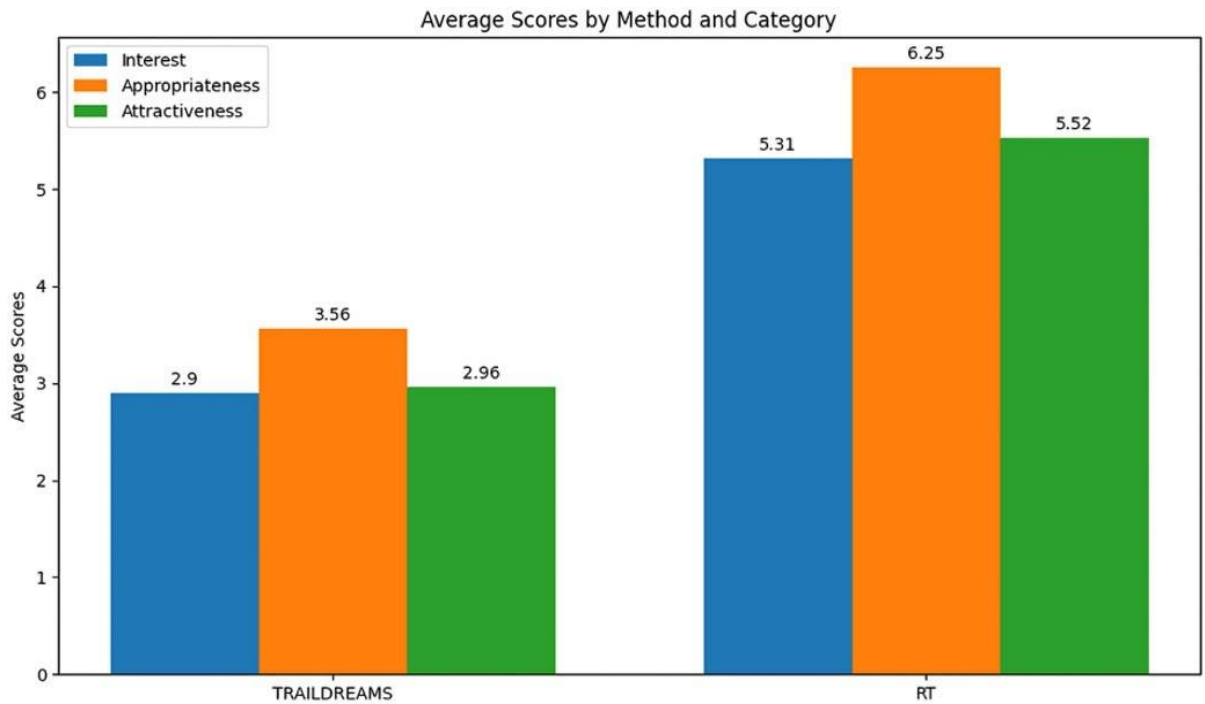

*Figure 8. TRAILDREAMS framework's trailers vs. real trailers scores (Source: Authors' own elaboration)*

## ANALYSIS AND FUTURE WORK

TRAILDREAMS framework marks a notable step forward in the field of automated trailer generation, showcasing considerable potential. However, to enhance its functionality and make its output more comparable to human-crafted trailers, several improvements are needed. **Figure 8** illustrates the disparity in aggregated scores between TRAILDREAMS generated trailers and real trailers across three films, highlighting substantial differences that exist between automated and traditional trailer production methods.

We conducted an in-depth qualitative analysis to discern the main differences between our trailers and real-world trailers. This exploration helped us identify several key areas in need of improvement. For each identified issue, we have proposed a viable solution, aiming to enhance the overall quality of our trailers. **Standard Clip Refinement**

*Problem*

Currently, SC may include scenes where characters are speaking without the corresponding audio, leading to a mismatch between what viewers see and hear.

*Future work*

Integrating an action recognition deep learning model could help identify clips where characters are prominently speaking during close-up shots. This model would enable the system to avoid selecting these clips or to ensure that if a speaking scene is chosen, it includes the corresponding dialogue or voice-over. This enhancement will create a more coherent and engaging viewing experience by aligning audio-visual elements more accurately.

**Voice Generation Improvement**

*Problem*

The voice-over generated by the current TTS models sometimes lacks the expressiveness and natural quality of human speech.

*Future work*

As TTS technology evolves, incorporating more advanced models will be crucial. Future iterations of
TRAILDREAMS should explore the integration of next-generation TTS engines that offer more lifelike and expressive voice outputs. This upgrade will significantly enhance the auditory appeal of the trailers, making them more engaging and human-like.

### Music Generation Enhancement

*Problem*

The generated music sometimes does not align perfectly with the trailer's emotional tone or narrative flow and may not always be aesthetically pleasing.

*Future work*

The development of more advanced music generation models will allow for the creation of soundtracks that are more harmonious. This can lead to an overall better viewing experience.

### Image Retrieval Process Optimization

*Problem*

The image retrieval process sometimes selects frames that do not correspond to the intended scenes outlined by LLM during the subplot generation phase.

*Future work*

To improve the visual coherence of the trailers, enhancing the robustness of the image retrieval process is essential. A potential solution could involve employing a multi-modal LLM, or "VLLM" (video large language model), to perform a "second check" to confirm that the selected frames accurately portray the intended narrative segments.

### Incorporation of Sound Effects

*Problem*

Unlike human-made trailers, TRAILDREAMS currently does not include sound effects, which are crucial for creating a dynamic and immersive trailer experience.

*Future work*

To bridge this gap, future versions of TRAILDREAMS should include a module for integrating relevant sound effects throughout the trailer. This could involve developing an intelligent system capable of identifying moments within the trailer where sound effects could enhance the narrative or emotional impact. Implementing this feature would add a new layer of depth and realism to the trailers, aligning them more closely with traditional human-crafted trailers.

## CONCLUSION

In this study, we introduced TRAILDREAMS, an innovative framework utilizing LLMs to advance the field of automated trailer generation. TRAILDREAMS distinguishes itself by integrating a multimodal approach, orchestrating audio, video, voiceovers, and dialogues, thereby crafting trailers that resonate deeply with audiences, offering a narrative richness that surpasses current standards in automated trailer production. Our findings, underscored by a comparative analysis, demonstrate TRAILDREAMS' superiority in creating more engaging, appropriate, and attractive trailers compared to its competitors. This framework's efficiency, exemplified by the rapid generation of a comprehensive trailer for "The Wolverine," showcases its potential to significantly streamline the trailer production process.

Despite its advancements, TRAILDREAMS, like all current automated systems, still trails behind human creativity for high-budget productions, where the interplay of artistry and technical finesse defines the industry standard. However, TRAILDREAMS emerges as a particularly valuable asset for independent filmmakers and low-budget productions. In these contexts, where resources are scarce, TRAILDREAMS offers a viable option.

As we look toward future enhancements—improving voice generation, refining clip selection, optimizing image retrieval, and incorporating sound effects—TRAILDREAMS reduces the gap between automated and human-generated trailers further.

In conclusion, TRAILDREAMS sets a new benchmark in the field of automated trailer generation, demonstrating the potential of AI to augment and, in some cases, transform the creative processes within the film industry.

**Data availability:** The source code for this article, along with the various crafted prompts, is available in the "TRAILDREAMS-framework" (Balestri et al., 2024b) repository, which can be accessed at https://github.com/ robertobalestri/TRAILDREAMS-Framework. Due to copyright restrictions, direct access to the generated trailers for many movies cannot be provided. However, a trailer for the public domain movie Night of the Living Dead (1968) by George A. Romero can be viewed at https://youtu.be/O9fS8s2LRqM. Other generated trailers will be shared on reasonable request to the corresponding author.

# REFERENCES

Alberani, D. (2006). *Cinemagoer* [Computer software]. *GitHub*. https://cinemagoer.github.io/

Baek, Y., Lee, B., Han, D., Yun, S., & Lee, H. (2019). Character region awareness for text detection. In *Proceedings of the 2019 IEEE/CVF Conference on Computer Vision and Pattern Recognition* (pp. 9357–9366). IEEE. https://doi.org/10.1109/CVPR.2019.00959

Balestri, R., Cascarano, P., Degli Esposti, M., & Pescatore, G. (2024a). An automatic deep learning approach for trailer generation through large language models. In *2024 9th International Conference on Frontiers of Signal Processing (ICFSP), Paris, France* (pp. 93–100). https://doi.org/10.1109/ICFSP62546.2024.10785516

Balestri, R., Cascarano, P., Degli Esposti, M., & Pescatore, G. (2024b). *TRAILDREAMS-framework* [Computer software]. *GitHub*. https://github.com/robertobalestri/TRAILDREAMS-Framework

Bellard, F. (2000). *FFmpeg* [Computer software]. https://ffmpeg.org/

Brachmann, C., Chunpir, H. I., Gennies, S., Haller, B., Kehl, P., Mochtarram, A. P., Möhlmann, D., Schrumpf, C., Schultz, C., Stolper, B., Walther-Franks, B., Jacobs, A., Hermes, T., & Herzog, O. (2009). Automatic movie trailer generation based on semantic video patterns. In I. Maglogiannis, V. Plagianakos, & I. Vlahavas (Eds.), *Artificial intelligence: Theories and applications. SETN 2012. Lecture notes in computer science* (Vol. 7297, pp. 345–352). Springer. https://doi.org/10.1007/978-3-642-30448-4_44

Bredin, H., Yin, R., Coria, J. M., Gelly, G., Korshunov, P., Lavechin, M., Fustes, Di., Titeux, H., Bouaziz, W., & Gill, M. P. (2020). Pyannote.Audio: Neural building blocks for speaker diarization. In *Proceedings of the ICASSP 2020–2020 IEEE International Conference on Acoustics, Speech and Signal Processing* (pp. 7124–7128). IEEE. https://doi.org/10.1109/ICASSP40776.2020.9052974

Castellano, B. (2014). *PySceneDetect* [Computer software]. https://www.scenedetect.com/

Copet, J., Kreuk, F., Gat. Itai, Remez, T., Kant, D., Synnaeve, G., Adi, Y., & Défossez, A. (2024). Simple and controllable music generation. *arXiv*. https://doi.org/10.48550/arXiv.2306.05284   Coqui AI. (2021). *XTTS* [Computer software]. https://docs.coqui.ai/en/latest/models/xtts.html   De Palma, B. (Director). (1996). Mission: Impossible [Film]. *Paramount Pictures*.

Degli Esposti, M., & Pescatore, G. (2023). Exploring TV seriality and television studies through data-driven approaches. In *Proceedings of the 13th Media Mutations International Conference*. https://doi.org/10.21428/ 93b7ef64.ec022085

Epstein, Z., Hertzmann, A., Akten, M., Farid, H., Fjeld, J., Frank, M. R., Groh, M., Herman, L., Leach, N., Mahari, R., Pentland, A., Russakovsky, O., Schroeder, H., & Smith, A. (2023). Art and the science of generative AI. *Science, 380*(6650), 1110–1111. https://doi.org/10.1126/science.adh4451

Explosion AI. (2016). *spaCy English models* [Computer software]. https://spacy.io/models/en

Gallifant, J., Fiske, A., Levites Strekalova, Y. A., Osorio-Valencia, J. S., Parke, R., Mwavu, R., Martinez, N., Gichoya, J. W., Ghassemi, M., Demner-Fushman, D., McCoy, L. G., Celi, L. A., & Pierce, R. (2024). Peer review of GPT–4 technical report and systems card. *PLOS Digital Health, 3*(1), Article e0000417. https://doi.org/10.1371/journal.pdig.0000417

Hesham, M., Hani, B., Fouad, N., & Amer, E. (2018). Smart trailer: Automatic generation of movie trailer using only subtitles. In *Proceedings of the 1st International Workshop on Deep and Representation Learning* (pp. 26–30). https://doi.org/10.1109/IWDRL.2018.8358211

Hu, Y., Jin, L., & Jiang, X. (2022). A GCN-based framework for generating trailers. In *Proceedings of the 8th International Conference on Computing and Artificial Intelligence* (pp. 610–617). https://doi.org/10.1145/ 3532213.3532306

Irie, G., Satou, T., Kojima, A., Yamasaki, T., & Aizawa, K. (2010). Automatic trailer generation. In *Proceedings of the 18th ACM International Conference on Multimedia* (pp. 839–842). ACM. https://doi.org/10.1145/ 1873951.1874092

Jackson, P. (Director). (2012). The Hobbit: An unexpected journey [Film]. *Warner Bros*.

JaidedAI. (2023). EASYOCR [Computer software]. *GitHub*. https://github.com/JaidedAI/EASYOCR jianfch. (2023). Stable Whisper [Computer software]. *GitHub*. https://github.com/jianfch/stable-ts

Likert, R. (1932). A technique for the measurement of attitudes. *Archives of Psychology, 22*(140), Article 55.

Long, F., Qiu, Z., Yao, T., & Mei, T. (2024). VideoDrafter: Content-consistent multi-scene video generation with LLM. *arXiv*. https://doi.org/10.48550/arXiv.2401.01256

Mahasseni, B., Lam, M., & Todorovic, S. (2017). Unsupervised video summarization with adversarial LSTM networks. In *Proceedings of the 2017 IEEE Conference on Computer Vision and Pattern Recognition* (pp. 2982– 2991). IEEE. https://doi.org/10.1109/CVPR.2017.318

Marhon, S. A., Cameron, C. J. F., & Kremer, S. C. (2013). Recurrent neural networks. In M. Bianchini, M. Maggini, & L. Jain (Eds.), *Handbook on neural information processing. Intelligent systems reference library* (Vol. 49, pp. 29–65). Springer. https://doi.org/10.1007/978-3-642-36657-4_2   Nolan, C. (Director). (2014). Interstellar [Film]. *Paramount Pictures*.

Oliveira, D. (2024, January 7). Creating movie trailers with AI. *Towards AI*. https://towardsai.net/p/machinelearning/creating-movie-trailers-with-ai

OpenAI. (2023a). OpenAI–GPT-4. *OpenAI*. https://openai.com/gpt-4

OpenAI. (2023b). Whisper [Computer software]. *GitHub*. https://github.com/openai/whisper

Papalampidi, P., Keller, F., & Lapata, M. (2021). Film trailer generation via task decomposition. *arXiv*. https://doi.org/10.48550/arXiv.2111.08774

Pavel, A., Reed, C., Hartmann, B., & Agrawala, M. (2014). Video digests: A browsable, skimmable format for informational lecture videos. In *Proceedings of the 27th Annual ACM Symposium on User Interface Software and Technology* (pp. 573–582). ACM. https://doi.org/10.1145/2642918.2647400

Piccolomini, E. L., Gandolfi, S., Poluzzi, L., Tavasci, L., Cascarano, P., & Pascucci, A. (2019). Recurrent neural networks applied to GNSS time series for denoising and prediction. In *Proceedings of the 26th International Symposium on Temporal Representation and Reasoning*. https://doi.org/10.4230/LIPIcs.TIME.2019.10

Radford, A., Kim, J. W., Hallacy, C., Ramesh, A., Goh, G., Agarwal, S., Sastry, G., Askell, A., Mishkin, P., Clark, J., Krueger, G., & Sutskever, I. (2021). Learning transferable visual models from natural language supervision [Preprint]. *arXiv*. https://arxiv.org/abs/2103.00020

Radford, A., Kim, J. W., Hallacy, C., Ramesh, A., Goh, G., Agarwal, S., Sastry, G., Askell, A., Mishkin, P., Clark, J., Krueger, G., & Sutskever, I. (2022). *clip-ViT-L-14* [Computer software]. https://huggingface.co/sentencetransformers/clip-ViT-L-14

Ratcliff, J. W., & Metzener, D. (1988). Pattern matching: The gestalt approach. *Dr. Dobb's Journal, 13*, Article 46. Rehusevych, O., & Firman, T. (2020). movie2trailer: Unsupervised trailer generation using anomaly detection. In D. Tabernik, A. Lukezic, & K. Grm (Eds.), *Proceedings of the 25th Computer Vision Winter Workshop*.

Reimers, N., & Gurevych, I. (2019). Sentence-BERT: Sentence embeddings using Siamese BERT-networks. In *Proceedings of the EMNLP-IJCNLP 2019–2019 Conference on Empirical Methods in Natural Language Processing and the 9th International Joint Conference on Natural Language Processing*. https://doi.org/10.18653/v1/d19-1410

Richards, G. (2018, March 14). Going in deep: How have movie trailers changed in the last decade? *Exit6 Film Festival Blog*. https://www.exit6filmfestival.com/post/2018/03/14/going-in-deep-how-have-movietrailers-changed-in-the-last-decade

Rouard, S., Massa, F., & Défossez, A. (2023). Hybrid transformers for music source separation. In *ICASSP 2023 - 2023 IEEE International Conference on Acoustics, Speech and Signal Processing (ICASSP), Rhodes Island, Greece* (pp. 1–5). https://doi.org/10.1109/ICASSP49357.2023.10096956

Sentence Transformers. (2022). *clip-ViT-L-14*. https://huggingface.co/sentence-transformers/clip-ViT-L-14

Shi, B., Bai, X., & Yao, C. (2017). An end-to-end trainable neural network for image-based sequence recognition and its application to scene text recognition. *IEEE Transactions on Pattern Analysis and Machine Intelligence, 39*(11), 2298–2304. https://doi.org/10.1109/TPAMI.2016.2646371

Smeaton, A. F., Lehane, B., O'Connor, N. E., Brady, C., & Craig, G. (2006). Automatically selecting shots for action movie trailers. In *Proceedings of the 8th ACM International Workshop on Multimedia Information Retrieval* (pp. 231–238). ACM. https://doi.org/10.1145/1178677.1178709

Smith, J. R., Joshi, D., Huet, B., Hsu, W., & Cota, J. (2017). Harnessing A.I. for augmenting creativity: Application to movie trailer creation. In *Proceedings of the 25th ACM International Conference on Multimedia* (pp. 1799–1808). ACM. https://doi.org/10.1145/3123266.3127906

Snyder, B. (2005). *Save the cat!: The last book on screenwriting you'll ever need*. Michael Wiese Productions.

Tarwani, K. M., & Edem, S. (2017). Survey on recurrent neural network in natural language processing. *International Journal of Engineering Trends and Technology, 48*(6), 301–304. https://doi.org/10.14445/22315381/IJETT-V48P253

Wasko, J. (2003). *How Hollywood works*. SAGE. https://doi.org/10.4135/9781446220214

Xie, J., Chen, X., Zhang, T., Zhang, Y., Lu, S.-P., Cesar, P., & Yang, Y. (2023). Multimodal-based and aestheticguided narrative video summarization. *IEEE Transactions on Multimedia, 25*, 4894–4908. https://doi.org/10.1109/TMM.2022.3183394

Xu, H., Zhen, Y., & Zha, H. (2015). Trailer generation via a point process-based visual attractiveness model. In *Proceedings of the 24th International Conference on Artificial Intelligence (IJCAI'15)* (pp. 2198–2204). AAAI Press. https://dl.acm.org/doi/10.5555/2832415.2832554

Zen, H., Dang, V., Clark, R., Zhang, Y., Weiss, R. J., Jia, Y., Chen, Z., & Wu, Y. (2019). Libritts: A corpus derived from LibriSpeech for text-to-speech. In *Proceedings of the Annual Conference of the International Speech Communication Association*. https://doi.org/10.21437/Interspeech.2019-2441

Zhou, H., Hermans, T., Karandikar, A. V., & Rehg, J. M. (2010). Movie genre classification via scene categorization. In *Proceedings of the 18th ACM International Conference on Multimedia* (pp. 747–750). ACM. https://doi.org/10.1145/1873951.1874068

Zhou, K., Qiao, Y., & Xiang, T. (2018). Deep reinforcement learning for unsupervised video summarization with diversity-representativeness reward. In *Proceedings of the 32nd AAAI Conference on Artificial Intelligence*. AAAI. https://doi.org/10.1609/aaai.v32i1.12255

Zhu, J., Yang, H., He, H., Wang, W., Tuo, Z., Cheng, W.-H., Gao, L., Song, J., & Fu, J. (2023). MovieFactory: Automatic movie creation from text using large generative models for language and images. In *Proceedings of the 31st ACM International Conference on Multimedia* (pp. 9313–9319). ACM. https://doi.org/10.1145/3581783.3612707